Assessment of hydrodynamic characteristics and computational resources for submarine resistance analysis: A comparative study between CFD Codes with application of BB2 Submarine

Noh Zainal Abidin[1,2,3], Frederic Grondin[2], Pol Muller[3] and Jean-François Sigrist[4]

[1]Faculty of Defence Science and Technology, National Defence University of Malaysia, Malaysia
[2]Research Institute in Civil and Mechanical Engineering (GEM), Ecole Centrale de Nantes, France
[3]Modelling and Simulation in Hydrodynamics, Sirehna, Naval Group, France
[4]Naval expert, eye-pi, France
noh.bin-zainal-abidin@ec-nantes.fr (NZ Abidin), frederic.grondin@ec-nantes.fr (F Grondin)

## 1.0 Abstract

Submarines are vital for maritime defense, requiring optimized hydrodynamic performance to minimize resistance. Advancements in Computational Fluid Dynamics (CFD) enable accurate predictions of submarine hydrodynamics for optimal design. This study compared the meshing capabilities of OpenFOAM and commercial software as well as the performance of High-Performance Computing (HPC) and standard PC resources upon hydrodynamic characteristics. The RANS turbulence model with $k - \omega\ SST$ was employed to analyze the resistances of the MARIN's BB2-class submarine. CFD simulations were conducted at a model scale (1:35.1) at a speed of 1.8235 m/s ($U_s$ of 21 knots) upon various mesh densities from 1 to 97 million cells. Empirical equations were initialized for turbulence parameters. Mesh sensitivity and iteration convergence ensured validated results. The findings showed that the results were validated with errors ranging from 0.3% to 10% across different mesh densities. The lowest error (0.3%) was achieved with 97 million cells generated by the commercial meshing tool with HPC, while 13 million cells by OpenFOAM with a standard PC resulted in a 3.4% error. Accuracy improved with precise initialization of turbulence parameters, mesh strategy, numerical schemes, and computing resources. The application of a standard PC with the OpenFOAM meshing tool was able to produce an acceptable accuracy, with less than 5% error for lower mesh densities. Thus, it can be suggested that using a standard PC was beneficial for preliminary hydrodynamic simulations. However, HPC with commercial software was essential for detailed industrial analyses, such as full-scale resistance and propulsion simulations.

## 2.0 Introduction

In 1872, William Froude revolutionized maritime engineering by constructing the world's first physical basin for testing ship models. This innovation spurred the development of numerous tank facilities globally dedicated to examining various aspects of ship hydrodynamics, such as resistance, propulsion, maneuverability, and seakeeping. While more economical and quicker than full-scale measurements, these model tests paved the way for advancements in virtual testing through computer simulations. Numerical simulations offer distinct advantages due to their speed and ease of optimization, significantly reducing the time needed for conceptual design analysis to mere days (Gao et al., 2018). The 28th ITTC (2017) introduced comprehensive guidelines for applying CFD in ship analysis to address cost concerns further. CFD not only cuts expenses but also yields highly detailed data, enriching fluid dynamics insights. Selecting appropriate turbulence models, such as RANS, is crucial for accuracy in CFD simulations. Additionally, ensuring a properly set mesh with high cell density is vital for reliable and convergent results, although computational power limits the number of cells that can be used (Jasak et al., 2019). Recent advancements in High Performance Computing (HPC) have significantly enhanced CFD methods, allowing for higher grid densities, more parallel processors, and accelerated convergence times. Despite these benefits, the high costs associated with acquiring and maintaining HPC systems remain a significant challenge, particularly for smaller organizations and research institutions. The research explored CFD methodology by conducting numerical simulations on the SSK class attack submarine BB2 at a 1:35.1 model scale (Overpelt et al., 2015). MARIN provided the 3D CAD of the full-scale submarine, while Sirehna, Naval Group supplied lab-scale data and high mesh density for verification. This research used HPC resources of the GLiCID Computing Facility (Ligerien Group for Intensive Distributed Computing, https://doi.org/10.60487/glicid, Pays de la Loire, France) and processed with OpenFOAM 11. Mesh generation, ranging from 1 to 97 million cells, was done using SnappyHexMesh and the Cadence Fidelity meshing tool. The results of both approaches were validated with experimental and highest mesh density. A summary of the comparison between PC and HPC was also studied. Eventually, the local viscous and pressure forces along the hull were investigated to determine the spot that experienced the highest stress.

## 3.0 Numerical CFD setup
This study applied the finite volume method to discretize the computational domain of the Navier-Stokes (NS) equation into small control volumes. All the solution fields were stored in the centroid of the control volume. The segregated solver was utilized to solve the scalar matrix equations in an iterative sequence. The RANS turbulence model with wall function ($y^+ >30$) and wall resolved ($y^+ <5$) was utilized in this research. The numerical parameters were configured in the open-source CFD code, OpenFOAM11. In this study, we simulated the submarine in a fully submerged isovolume and isothermal in steady state condition to simplify the equation of NS equation as follows:

$$\nabla . (\boldsymbol{u}) = 0 \quad (1)$$

$$\frac{\partial (\boldsymbol{u})}{\partial t} + \nabla . (\boldsymbol{uu}) = \frac{-\nabla p}{\rho} + \nu \nabla^2 (\boldsymbol{u}) \quad (2)$$

The closure equations, typically derived from experimental data and adjusted for optimal performance across various flow conditions (Wilcox, 2008). The initialization of $k$ and $\omega$ is significance to ensure the solution reach convergence and avoid numerical diffusion. In this study, the value of $k$ and $\omega$ can be initialized approximately according to viscosity ratio, $\frac{\mu_t}{\mu}$ and turbulence intensity, $I$ as follows:

$$\omega_{far} = \frac{\rho k}{\mu}\left(\frac{\mu_t}{\mu}\right)^{-1} , \quad k_{far} = \frac{3}{2}(UI)^2 \quad (3)$$

Since this case was incompressible and in a steady-state condition, the transient term was set to be steady state. The RANS of $k-\omega\ SST$ was utilized as a turbulence model, which is capable of solving for both low and high $y^+$ values (Menter et al., 2003). The solution solver employed was the Semi-Implicit Method for Pressure-Linked Equations with Consistence (SIMPLEC). For the initial stage of verification and validation, the free stream velocity, U, utilized was 21 knots at real scale, as referred to by Overpelt et al. (2015), and was scaled down by Froude similitude as recommended by ITTC (2014) at $U_m = U_s/\sqrt{\lambda}$ with a ratio of (1:35.1) to ensure similarity with the model test (Lab-scale) provided by Sirehna, Naval Group. The BB2 submarine particulars were as shown in Table 1, while the initialization of physics and turbulence parameters was as shown in Table 2.

Table 1. Submarine BB2 particulars

| Description | Symbol / unit | Full (1:1) | Model (1:35.1) |
|---|---|---|---|
| Length overall | L (m) | 70.2 | 2 |
| Beam | B (m) | 9.6 | 0.2735 |
| Depth (to deck) | D(m) | 10.6 | 0.3020 |
| Depth (to top of sail) | Dsail(m) | 16.2 | 0.4615 |

Table 2. Initialization physic and turbulence

| Parameter | Unit | Value |
|---|---|---|
| $U_m$ | $m/s$ | 1.8235 |
| $k$ | $m^2/s^2$ | 0.0005 |
| $\omega$ | $1/s$ | 498.77 |
| $\nu_w$ | $m^2/s$ | 1e-06 |
| $\rho$ | $kg/m^3$ | 1000 |
| $\mu_t/\mu$ | - | 1 |
| $I$ | - | 1% |
| $Re$ | - | 3.65e06 |

The submarine simulation is in fully submerged domain as constructed the computational domain w.r.t to L= 2m as (8m x 14m x 8m) referred to 28th ITTC (2017) that is affordable to simulate the flow efficiently as shown Fig.1. An adequate computational domain ensures numerical stabilities, avoiding the backflow and numerical oscillations may arise that affecting the accuracy and reliability of the simulation results.

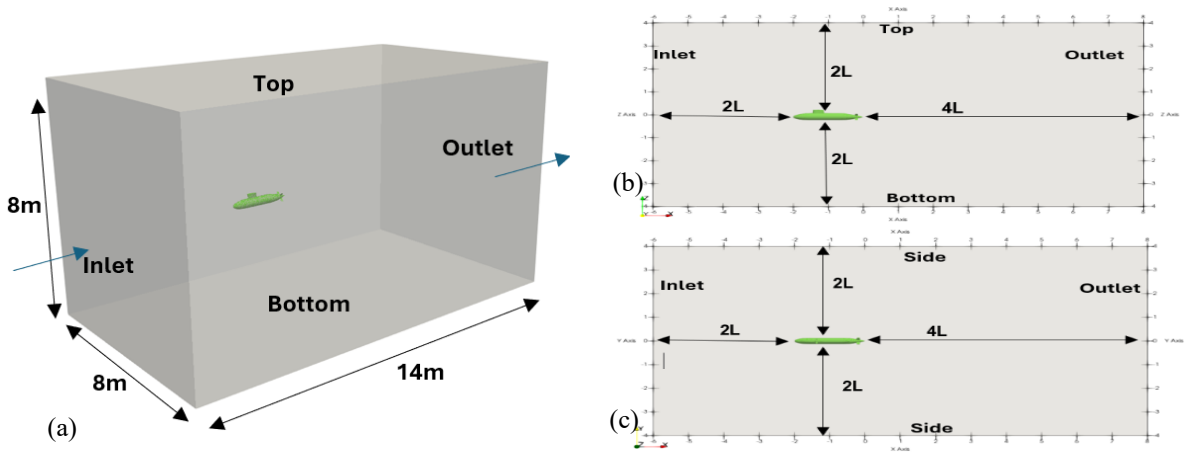

Fig.1: (a) 3D Computational Domain and Boundary Conditions, (b) Side view, (c) Top View

The near-wall treatment on the submarine hull was set based on the $y^+$ value. In this study, the mesh was generated in 3 cases, including (*i*) mesh from 1 to 13 million cells, which was generated by snappyHexMesh in OpenFOAM using a standard PC, (*ii*) mesh of more than 20 million cells, which was generated by snappyHexMesh in OpenFOAM using HPC, and (*iii*) mesh of more than 20 million cells, which was generated by Cadence using HPC. For (*i*), initially, a hexahedral background mesh was created by the blockMesh utility to set up the global base size, $\Delta$ (dx, dy, dz), at the ratio of $\Delta/L$ =0.05. The ratio could be modified to refine or reduce the mesh density globally throughout the domain by $\Delta/\sqrt{2}$ based on ITTC (2017) and (Paredes et al., 2021), from low to high mesh densities. Then, snappyHexMesh was utilized to produce the castellated mesh, surface mesh snapping, and to add boundary layer configuration. The targeted boundary layer was 6 to 8 layers, as suggested by prior studies conducted by Paredes et al. (2021) and Jasak et al. (2019). Refinement on mesh surfaces with levels (5,6) was used to achieve a smooth surface and curvature. The refinement region tool was employed for two regions near the submarine wall, in which each region reduced the size of the cells, $\Delta/2^n$ as they approached closer to the submarine's hull, as presented in Fig. 2. The size of the elements in the inner refinement region, $\Delta_{ir}$ was adjusted according to the targeted $y^+$ value. The stretching factor, $r$, of 1.2 for boudary layer generation was utilized. The targeted first layer thickness, $y_1$, boundary layer thickness, $\delta$, and number of prism layers, $m$, could be approximated based on the targeted $y^+$ (White & Majdalani, 2022), as represented in the equation below.

$$\frac{y^+}{y_1} = \frac{0.0487 \cdot U_m}{v \cdot \ln(0.06 \cdot Re)} \; ; \quad \delta = \frac{0.16L}{Re^{1/7}} \; ; \quad m = \frac{\ln\left(1 - (1-r)\frac{\delta}{y_1}\right)}{\ln(r)} \quad (4)$$

In Fig. 2, (i) indicates the inner refinement region, $\Delta_{ir} = \Delta_{or}/2^n$, while (o) shows an outer region, $\Delta_{or} = \Delta_b/2^n$ and (b) background region, $\Delta$. The same strategy is utilized for case (*iii*) generated by Cadence for high mesh density.

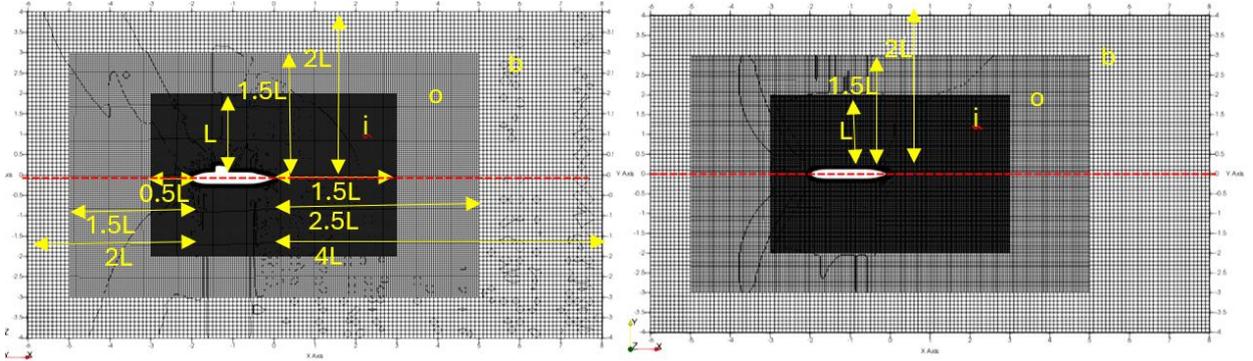

Fig. 2: Meshing Strategy, refinement zone dimension (Side and Top View). 13,705,468 cells (snappyHexMesh)

All the meshes generated for each case were processed in CFD code of OpenFOAM 11.The SIMPLEC with under relaxation factors on parameters $U$=0.9, $k$ and $\omega$ are 0.7 were utilized. For the gradient, divergence and Laplacian schemes were discretized using the second-order scheme: cellLimited Gauss linear 1, bounded Gauss linearUpwindV, and Gauss linear corrected. The simulation has been monitored up to 1000 iterations for convergence. Two iterations of non-orthogonal mesh correctors have been utilized to have stable solution. The geometric-algebraic multigrid solver (GAMG) with Gauss Seidel for smoother solver has been used for quantities of the transport equation. By varying the mesh density via mesh refinement factor systematically according to the targeted $y^+$, mesh convergence study has been conducted. Note that, by applying the mesh refinement, the numeric setup remains the same for all the cases as illustrated in Table 3.

## 4.0 Results and Discussion - Submarine hydrodynamic assessment

The HPC resources were employed for computations involving high mesh density, while a standard Pc was used for simulations with lower mesh density. The specifications of the resource utilized are shown in Table 3. While the summary of sample cases conducted is presented in Table 4. The global resistance of submarine is taken for studying the hydrodynamic assessment. The global resistance, $F_t$ contributed by integration of two components along the submarine which are pressure, $F_p$ and viscous, $F_v$ forces that solve in x-direction. Then the $F_t$ can be compared it with experimental results for verification and validation process (Overpelt et al., 2015). Initially, the mesh convergence study was conducted on Case 1 to 7 (1 – 13) million using snappyHexMesh with 6 cores of CPU. Then, the mesh for Case 5, 6 and 7 taken as a base for determining the order of convergence, $p$, estimated extrapolated solution, relative error, $e_{21}$ and grid convergence index (GCI) computed based on Richardson extrapolation referred to prior study (Celik et al., 2008) and ITTC guidelines (ITTC, 2017). It is shown that, the average $y^+$ reduces from 478 till 7.5 when increase the mesh density as shown in Fig. 3 and Table 6. Thus, the wall function is utilized for wall treatment in viscous sub layer region. However, the Avg. $y^+$ less than 10 for mesh of Case 5 -7 can be acceptable as referred from Paredes et al. (2021), which used a mesh of Avg. $y^+$= 16.48 for his submarine computation.

Table 3. Computing resources

| Resources | Type | Computing |
|---|---|---|
| HPC | Nautilus cluster in Glicid, 5376 AMD Genoa cores, 28TB RAM, 16 A100-80GB GPUs, 8 A40 GPUs, 100GB infiniband network | High Mesh Density (>20 millions) |
| Pc | Dell, 13th Gen Intel(R) Core(TM) i7-13850HX, 2100 Mhz, 20 Core(s), 32GB RAM | Lower Mesh Density (<20 millions) |

Table 4. Sample Cases

| Case | Mesh tool | Resources | Target $y^+$ |
|---|---|---|---|
| 1 - 7 | SnappyHex Mesh | Standard PC | 5 |
| 8 -9 | SnappyHexMesh | HPC | 1 |
| 10 - 13 | Cadence | HPC | 1 |

Fig.4 presents Case 1 – 7 simulation results using standard Pc. The trends are significant in providing information on mesh and iterative convergence concerning the quantity of interest (Qoi), $F_p$ and $F_v$ getting constant at specific iterations when mesh densities increase (Abidin et al., 2021). The results of the mesh sensitivity study for Mesh Cases 5 to 7 are shown in Table 6.

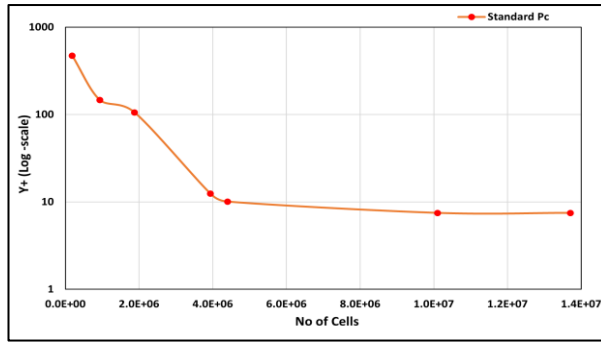 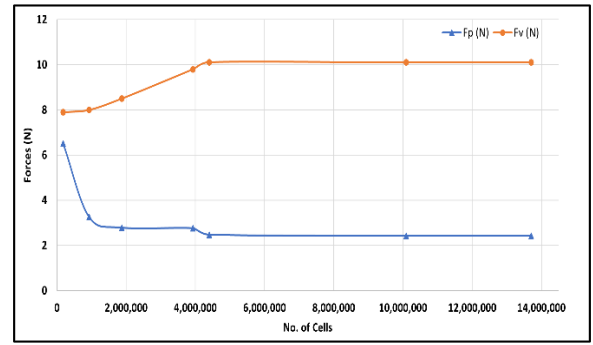

Fig.3: The average Log-Scale ($y^+$) for Cells (180k – 13 million)    Fig.4: The prediction $F_p$ and $F_v$

The results of the mesh sensitivity study and Richardson extrapolation procedure are shown in Table 5. It shown that the fine mesh of Case 7 is well converged in terms of the relative error, $e_{21}$ and the $CGI_{21}$. The parameter of $s$=1 indicates the monotonic convergence, as presented by Eça & Hoekstra (2014). While the refinement ratio, $r_{21}$ and $r_{32}$ are both greater than 1.1 indicating the meshes are sufficient for good mesh sensitivity study. As an error of 0.44% is likely to be acceptable for most engineering studies, the solution on an infinitely fine mesh can be used with confidence. Furthermore, the order of convergence, $p$ = 2, is the ideal order of convergence. Therefore, the extrapolated global resistance approximates as $F_0 \approx F_t$=12.48N. Hence, any further mesh refinement is likely to give noticeable improvements in the accuracy of global resistance and could be worthwhile if the computational cost is reasonable.

Table 5. Results of Mesh Convergence Study

| $r_{21}$(-) | $r_{32}$(-) | $s$ (-) | $e_{21}$(%) | $e_{21}^{ext}$(%) | $p$(-) | $CGI_{21}$(%) | $F_0$ (N) |
|---|---|---|---|---|---|---|---|
| 1.107 | 1.369 | 1 | 0.08 | 0.35% | 2 | 0.44% | 12.4758 |

Case 7 illustrated good CGI. Nevertheless, the verification and validation can be achieved by comparing them with experimental data from Sirehna, Naval Group. The summary of errors and computation times obtained for the entire study is presented in Table 6. Cases 8 and 9 generated the high mesh density using snappyHexMesh via HPC. While for cases 10 to 13, the high mesh density generated by Cadence can target the maximum Avg. $y^+$ less than 1. Thus, wall resolved was applied to compute the viscous sub-layer accurately. It can be observed that Cases 4 to 7 (standard Pc) and 10 to 13 (HPC) illustrated errors of less than 5%. A comparison between the accuracy, computing times and meshing tool for standard Pc and HPC is presented in Fig. 5. As discussed earlier, the mesh of Case 7 obtained good CGI and produced error of 3.4% using the OpenFOAM meshing tool. However, in Cases 8 and 9, the error increased to 9.4%, and the accuracy did not improve. This was due to poor generation and collapsing of cells vertex onto surface in the boundary layer zone, which negatively affected the interpolation of solution.While for Case 10 to 13, with high mesh density presented accuracy improve until the error reduces to 0.3%. Fig. 5 (right) illustrates the comparison between accuracy and computation on both resources utilized. It can be observed the computation time of Case 7 (13 million) had overtaken by Case 10 (26 million) due to contribution number of cores in HPC. The time spent increases with increment of mesh density. In addition, Table 6 and Fig. 5 presented the comparison on surface mesh generation between both tools. We discovered the OpenFOAM meshing tool has difficulty on generating the targeted boundary layer of $y^+$ in contrast to Cadence.

Table 6. Summary of Simulation on Various Mesh Densities

| Case | Cells No. | CPU | Station | Tool | Wall/No | yplus Min | yplus Max | yplus Avg | Ftcfd | Error (%) | Clock (s) | CPU (h) |
|---|---|---|---|---|---|---|---|---|---|---|---|---|
| 1 | 180,348 | 6 | Pc | OF | Wall | 3.941 | 471.464 | 474 | 14.41 | 19.07 | 84.00 | 0.14 |
| 2 | 929,873 | 6 | Pc | OF | Wall | 1.10591 | 189.734 | 147 | 11.26 | 6.97 | 489.00 | 0.82 |
| 3 | 1,873,262 | 6 | Pc | OF | Wall | 0.28105 | 153.16 | 106 | 11.29 | 6.75 | 1143.70 | 1.91 |
| 4 | 3,932,029 | 6 | Pc | OF | Wall | 0.12336 | 131.199 | 12.5 | 12.59 | 4.00 | 2475.00 | 4.13 |
| 5 | 4,400,222 | 6 | Pc | OF | Wall | 0.13336 | 120.059 | 10.1 | 12.56 | 3.70 | 2950.00 | 4.92 |
| 6 | 10,092,709 | 6 | Pc | OF | Wall | 0.13369 | 113.702 | 7.51278 | 12.53 | 3.52 | 7314.00 | 12.19 |
| 7 | 13,705,468 | 6 | Pc | OF | Wall | 0.1086 | 113.713 | 7.51383 | 12.52 | 3.44 | 10118.00 | 16.86 |
| 8 | 38,185,848 | 50 | HPC | OF | No | 0.00199 | 88.3181 | 1.06215 | 10.95 | 9.51 | 6492.44 | 90.17 |
| 9 | 39,233,777 | 50 | HPC | OF | No | 0.00228 | 96.4607 | 1.22428 | 10.96 | 9.42 | 6509.73 | 90.41 |
| 10 | 26,724,343 | 50 | HPC | Cadence | No | 0.005 | 4.50451 | 0.85224 | 12.49 | 3.22 | 4300.38 | 59.73 |
| 11 | 50,071,757 | 50 | HPC | Cadence | No | 0.01051 | 4.50564 | 0.95853 | 11.78 | 2.68 | 8631.01 | 119.88 |
| 12 | 88,115,356 | 50 | HPC | Cadence | No | 0.00378 | 1.7943 | 0.22413 | 12.19 | 0.69 | 16479.10 | 228.88 |
| 13 | 97,092,477 | 50 | HPC NG | Cadence | No | 0.00244 | 0.77986 | 0.24248 | 12.13 | 0.24 | 18358.40 | 254.98 |

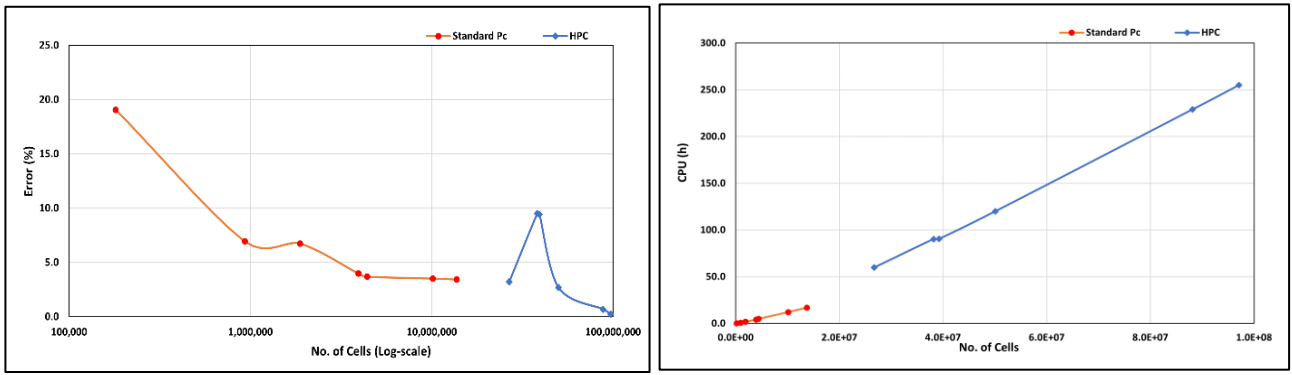

Fig.5: The accuracy and CPU time for standard PC (6 cores) and HPC (50 cores)

Detailed observation on sail planes or rudder planes showed an unsmooth surface at the edges that was expected to influence the accuracy of wall shear stress as shown in Fig. 6. Fig. 7 illustrated the hydrodynamic forces of $F_v$ and $F_p$ distributed along the hull for each section from bow to tail of submarine. It was observed the $F_v$ is dominated over $F_p$ contributing as 85% towards global resistance. The $F_v$ is highest at the sail ($x/L = 0.8 – 0.5$) and tail ($x/L = 0.1 – 0$) regions. The turbulent field was substantially altered by the sail, sail planes, and rudder planes, which significantly modified the flow and influenced the distribution of $F_v$. The wake was influenced by the shear layer from the sail planes of fins' trailing edge as shown in Fig 8 (at straight ahead speed, $U_m = 1.8235$). The sail on the upper side of the hull generates an adverse pressure gradient, which results in the formation of a horseshoe vortex that travels downstream and interacts with the boundary layer of the hull. The sail on the upper side of the hull generated an adverse pressure gradient, which resulted in the formation of a horseshoe vortex that traveled downstream and interacted with the boundary layer of the hull. The sail's trailing edge and rudder section also generated additional vortices, as shown in similar studies by Rocca et al., (2022). Nevertheless, the flow structure was unable to clearly visualize the vortices generated accurately due to the steady solver and RANS assumption utilized in this study.

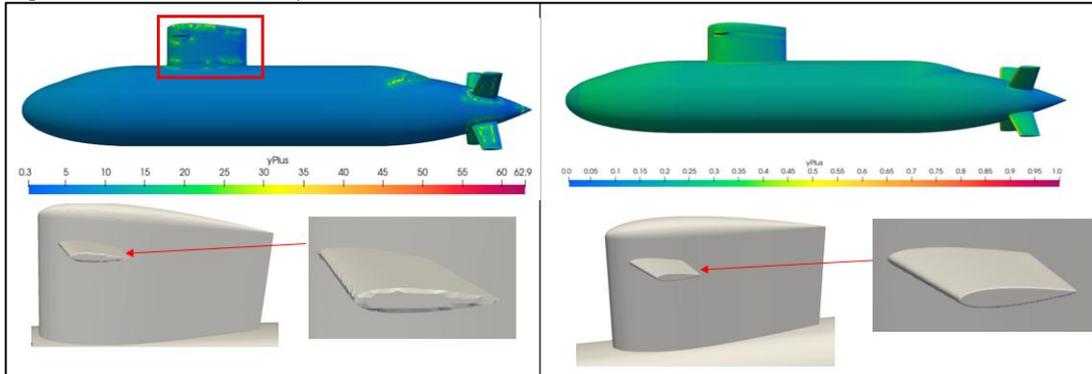

Fig.6: The comparison between meshing tools of SnappyHexMesh (Case 7) and Cadence (Case 11)

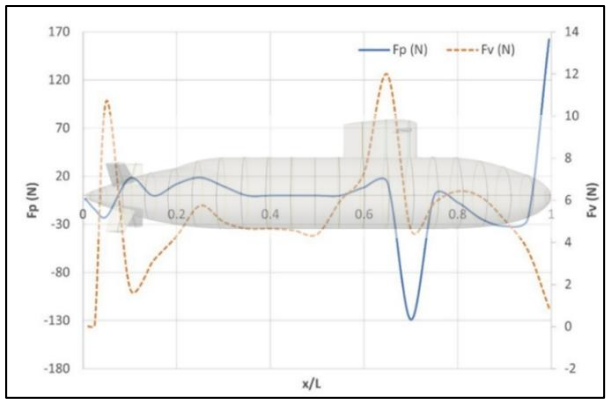 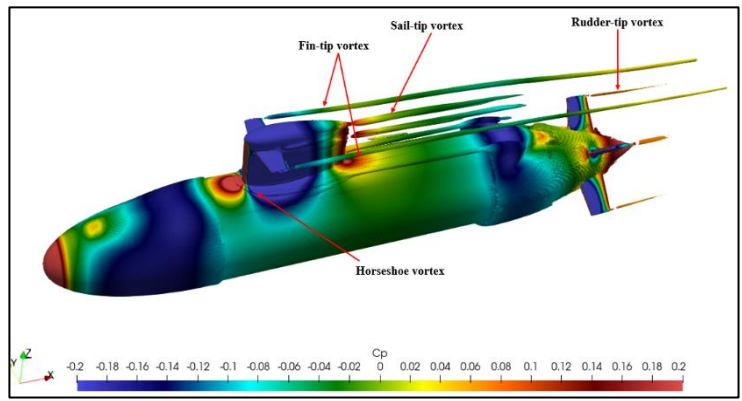

Fig.7: The Local Forces Distribution  Fig.8: The Flow characteristic along the hull (Q criterion)

**5.0 Conclusion**

A good agreement is observed between the CFD predictions and experimental data, particularly with the high density of mesh, regarding the hydrodynamic characteristics of the submarine hull with integral of $F_v$ and $F_p$ at various mesh densities. It is concluded that the mesh generated by OpenFOAM with standard Pc is acceptable, with an error less than 5%. Case 7 exhibiting a 3.4% error with GCI of 0.44% demonstrates an acceptable value and approximate by Richardson extrapolation method according to ITTC standards. However, the OpenFOAM meshing tool appeared to be unable to improve the accuracy of solution even though able to generate high density mesh such Case 8 and 9 as in Fig. 5. While the utilization of HPC shows advantages like flexibility in generating high-density mesh and allows for larger computation times. Moreover, the usage of commercial meshing tools such as Cadence is shown to accurately generate boundary layers according to targeted $y^+$ based on Eq.4, leading to an increase in global mesh density and a reduction of error to 0.3%. Rigorous assessment on Cases 1 to 7 demonstrates that standard Pc is significance for conducting preliminary ship hydrodynamic simulations while HPC systems are indispensable for handling more intricate and detailed industrial analyses, such as full-scale resistance and propulsion simulations. However, results obtained from standard Pc should be verified against high mesh density results (Case 13) or experiment before being utilized for further analysis. However, it is imperative to ensure correct initialization and boundary condition values, as discussed in the previous section. The choice of numerical scheme for discretizing transport equations significantly influences stability, accuracy, and solution boundedness based on specific cases. The investigation on global and local contributions of hydrodynamic components shows that the sail and rudder experience highest $F_v$ due to strong turbulent field, steep pressure gradients, and high velocity gradients. Our findings highlight the complementary roles of HPC and standard Pc, as well as the utilization of both commercial and open-source meshing tools, in optimizing the overall accuracy and cost-effectiveness of CFD simulations for ship hydrodynamics.